# Spectral-temporal processing using integrated recursive electro-optic circuit

Xudong Li[1,*,†], Yaowen Hu[2,*,†], Tong Ge[1,*], Andrea Cordaro[1,*], Yunxiang Song[1,3], Xinrui Zhu[1], Shengyuan Lu[1], Keith Powell[1], Letícia Magalhães[1], Urban Senica[1], Neil Sinclair[1], Marko Lončar[1,†]

[1] John A. Paulson School of Engineering and Applied Sciences, Harvard University, Cambridge, MA 02138, USA
[2] State Key Laboratory for Mesoscopic Physics and Frontiers Science Center for Nano-optoelectronics, School of Physics, Peking University, Beijing 100871, China
[3] Quantum Science and Engineering, Harvard University, Cambridge, MA 02139, USA

[*] *These authors contributed equally*
[†] xudongli@g.harvard.edu, yaowenhu@pku.edu.cn, loncar@g.harvard.edu

**Advances in integrated photonics have enabled unprecedented level of control of light, powering a wide range of photonic technologies from communications and computing to precision metrology and quantum information. However, the conventional on-chip optical signal processing approaches based on optical waveguides and cavities suffer from their large physical footprint and narrow operating bandwidth, respectively. To address this, we propose and experimentally demonstrate, using the thin film lithium niobate (TFLN) photonics, a modular, recursive optical signal processing framework. In our approach, fast electro-optic (EO) switch is used to either keep an optical packet inside the loop, with the processing element embedded within, or to release it from the loop and direct it towards the output waveguide. By configuring the switch on a timescale shorter than a time the photon spends traversing the loop, this architecture can achieve different optical pathlengths in a compact footprint without sacrificing optical bandwidth. As an example, we embed a phase-modulator (PM) inside the loop and demonstrate a frequency shift of optical packets up to 420 GHz, using only a 3 GHz sinusoidal microwave signal. By replacing the PM with chirped Bragg gratings (CBG), we demonstrate recursive delay line featuring large group delays of 28 ps/nm over a 30 nm optical bandwidth. Finally, by introducing asymmetric Mach-Zehnder interferometer (AMZI) inside the loop, we demonstrate a reconfigurable differentiation of the optical packet in time, up to an unprecedented fifth-order. Our results establish a powerful and scalable platform for multifunctional photonic processing, setting the stage for next-generation integrated systems with ultrahigh reconfigurability and spectral–temporal versatility.**

## Introduction

Nearly all applications of integrated photonics rely on efficient control of spectral and temporal properties of light fields. For example, coherent optical communications and microwave photonics protocols require precise control of the amplitude and phase of light[1,2], optical ranging crucially depends on interference between frequency-chirped forward-propagating and reflected optical signals[3–5], while emerging optical computing tasks leverage the efficient encoding of data on multiple light paths[6–9]. Advances in integrated photonics have enabled these and other applications through realization of low-insertion loss waveguides[10–12] and high-bandwidth, low-voltage, electro-optic (EO) modulators[13–15]. When combined with low-loss resonators, they have enabled broad EO frequency comb sources[16–18] and frequency shifters[19]. Applications that seamlessly integrate these components such as optical neural networks[20–22], line-by-line pulse shaping[23], femtosecond pulse generation[24], and microwave signal processing[25] have been demonstrated.

However, to unlock the full potential of integrated optical systems, the trade-off between optical processing bandwidth and device footprint needs to be overcome. While waveguide-based devices like modulators, $\chi^{(2)}$ or $\chi^{(3)}$ nonlinear frequency converters, and dispersive gratings can manipulate broadband optical signals, they usually need a long propagation length to achieve the desired effect (e.g. large phase shift, conversion efficiency and dispersion, respectively), resulting in large footprints. On the other hand, resonator-based devices are compact and benefit from enhanced light-matter interaction due to build-up of optical field inside the resonator, but only narrow bandwidth optical signals satisfy the resonance condition. This effectively restricts the input signal to be either continuous-wave (CW) or a frequency comb spaced by the cavity free spectral range (FSR), limiting their application space.

Here, we overcome these limitations by proposing and demonstrating a recursive processing framework that combines the large bandwidth of optical waveguides with the compact footprint of optical resonators. Conventionally, an optical pulse goes through several blocks ($O_1, O_2, .... O_N$ in Fig. 1a, left panel) daisy-chained on the same chip to perform desired processing task. An example is the optical finite-impulse response (FIR) filter[26,27] which can be built from successive Mach-Zehnder interferometers (MZIs) and delay lines. With our methods, the same processing task could be accomplished by keeping the pulse in a loop that consists of a reconfigurable block that can perform different functionalities at different points in time (Fig. 1a, right panel). While similar architectures have been implemented using fiber-optics to demonstrate recurrent neural networks[28] and Ising machines[29], the advancements of low loss thin-film lithium niobate (TFLN)[12] featuring fast on-chip modulators[13,14], allows the implementation of the whole system on a single chip (Fig. 1b). The recursive unit is essentially an optical resonator with its coupling rate to the feed-waveguide controlled by a MZI switch (also known as MZI-assisted ring[30–32]) that can be reconfigured on a time-scale shorter than the photon cavity roundtrip time. When the switch is open ("couple" state), it effectively serves as a directional coupler (with 100% coupling) that directs the pulse towards the optical block of interest (modulators, filters, optical gain, non-linear waveguides, etc.), and then couples it back to the output waveguide. When the switch is in closed ("circulate") state, the operation block is decoupled from the input/output waveguide, and the pulse is trapped inside the loop thus passing multiple times through the block.

While this approach allows for a large range of time and frequency domain operations to be implemented recursively (Fig. 1c), in this work we experimentally demonstrate three functionalities that conventionally require large footprints (Fig. 1a, left panel). First, we embed a phase modulator (PM) inside the recursive loop to demonstrate spectral shearing of 100 ps long (~20 GHz bandwidth) optical pulses up to 400 GHz, using a low-frequency control RF signal (3 GHz). This could be of interest for channel-switching in e.g. dense wavelength division multiplexing (DWDM) applications[33] (>10 channels), terahertz synthesis and detection[14] and quantum communications[34]. Next, we demonstrate a large on-chip dispersion of >25 ps/nm across a 30 nm wide bandwidth using a resonator based on chirped Bragg gratings (CBG) mirrors. This is of interest for generation of ultra short optical pulses[24], frequency domain computation[35], optical pulse shaping[36] and so on. Finally, by inserting an asymmetric Mach-Zehnder interferometer (AMZI) into our circuit, we perform differentiation of the optical pulse from the $0^{th}$ up to the $5^{th}$ order and generate pulse trains of different differentiation orders with arbitrary relative amplitudes by dynamically modulating the cavity coupling rate. This programmable differentiator can be used in photonic image processing[37] for instance.

## Results

**Characterization of MZI-based dynamically coupled resonator**

To demonstrate the operation of the switch-coupled resonator, continuous-wave (CW) light is injected into the device, and optical transmission spectrum is measured for different voltages applied to the switch (Fig. 2a). Strong suppression of the resonant peaks are observed when the switch is configured in the coupled and circulating states, corresponding to an over-coupled and under-coupled resonator, respectively. We also show that critical coupling can be achieved by tuning the bias voltage applied to the MZI switch.

Next, we demonstrate optical pulse trapping. A 100-ps pulse, generated by carving CW light by an off-chip amplitude modulator (AM), is sent into the circuit. By applying square wave voltage signals with different duty cycles to the MZI switch, we trap the pulse for up to 14 round trips (Fig. 2b). By fitting the peak output pulse power with an exponential decay curve (Fig. 2b inset), we evaluate the round-trip loss to be 0.96 dB/roundtrip, corresponding to an intrinsic quality factor $Q_i$ of around $1.3 \times 10^6$ and a photon lifetime (i.e. 3dB storage time) of around 1 ns.

**Multi-pass phase modulator for enhanced spectral shearing**

Next, we utilize our architecture to demonstrate EO frequency shifting of optical pulses with unprecedented shift-to-bandwidth ratio. When an optical wave packet $f(t)e^{i(\omega t - \beta z)}$ propagates through a PM with a linear voltage ramp applied to it, it acquires a phase $\phi(t) = \Delta f \cdot t$ and experiences an effective frequency shift of $\Delta f$ from the center frequency $\omega$ - a process called spectral shearing. For experimental simplicity, linear voltage ramps are often approximated with a linear portion of a sine wave[38–43]. Here, the rising or falling voltages of a sinusoidal drive signal with frequency $f_{RF}$ and amplitude $V$ is applied to PM with half-wave voltage $V_\pi$, resulting in a frequency shift $\Delta f = \pi \frac{V}{V_\pi} f_{RF}$. Therefore, large $\Delta f$ requires large driving frequency $f_{RF}$ and thus high-speed electronics, which can be challenging. More importantly, $f_{RF}$ puts an upper limit on the temporal width of the optical pulse that can be sheared since the pulse duration is required to be within the duration of the rising or falling slope of the sine wave, to avoid pulse distortion[44]. As a result, previous spectral shearing demonstrations fall short of generating large frequency shifting for long pulses, that is, have modest figure of merit $F = \frac{\Delta f}{FWHM}$, where $FWHM$ is the pulse's frequency bandwidth.

We overcome this limit by placing PM inside the recursive unit (Fig. 3a, 3b) and applying microwave control (Fig. 3c) to both the MZI switch and the PM to accomplish multi-round spectral shearing. We demonstrate $\Delta f = 420$ GHz for $FWHM = 20$ GHZ pulse trapped inside the loop and interacting with PM 14 times, resulting in $F > 20$ (Fig. 3d). Importantly, this large shift is achieved using a low frequency drive signal with $f_{RF} = 3$ GHz, that is 140 times smaller than the total shift achieved. Details of the experiment setup and control signals can be found in supplementary information. The pulse also experiences distortions, visible in Fig. 3d. The smaller peaks before and after the main peak are attributed to the leakages from an earlier and later roundtrips, respectively. We note that these leakages are more prominent than in Fig. 2b (without the shearing signal), suggesting that the strong RF drive may cause changes in the coupling condition, possibly due to heating of the chip. However, since these additional peaks are separated in both frequency and time from the main pulse, they can be easily filtered out by either band-pass filters or another MZI switch. Another distortion that we observe is the broadening of the sheared peak, especially for the pulses that undergo large number of shearing rounds. This is attributed to the small but finite difference between the round trip time and the period of the shearing signal: as a result the pulse eventually drifts away from the center of the rising slope of the shearing signal, causing it to interact with the non-linear part of the sinusoidal wave[44]. This could be mitigated by using either shorter pulses or lower $f_{RF}$. For quantitative simulation of the pulse deformation, see supplementary information.

By controlling the phase of the sinusoidal wave, the optical pulse is synchronized with the falling edge of the sinewave, resulting in the red shift. Fig. 3e shows the shearing results with >300 GHz blue- and red-detuned spectral shifts achieved with reduced RF driving power to minimize the heat-induced pulse distortion. Finally, a comparison between our results and previous work on spectral shearing is shown in Fig. 3f. Our recursive circuit approach, which allows an optical pulse to pass N times through PM, improves both figures of merit $\frac{\Delta f}{f_{RF}}$ and $\frac{\Delta f}{FWHM}$, bringing new possibilities to applications for both short and long pulses.

**Time-varying chirped Bragg grating cavity for large on-chip dispersion**

A passive element that can greatly benefit from recursive operations is the chirped Bragg grating (CBG). Inside the CBG different wavelengths of light are reflected at different locations, thus acquiring different time delays[45]. This is of interest for applications like pulse compression and signal processing, where long linearly chirped gratings are required to induce large linear dispersion. While the latter is challenging to realize on x-cut TFLN (due to strong birefringence), current cascaded straight CBGs [46,47] are limited to group delays <10 ps/nm, over a bandwidth on the order of tens of nanometers, despite large on-chip footprint they occupy. Here, we overcome this limitation using the recursive unit shown in Fig. 4a. We modify the circuit structure by terminating two sides of the MZI switch with CBGs, serving as mirrors. In this way, when the switch is configured into circulating state, the bottom part forms a linear Fabry-Perot (FP) cavity. Then, once a pulse is introduced into the FP cavity, it reflects from each CBG N times, each time acquiring additional delay. In the end, the MZI switch is configured into coupled state, and the pulse is released and directed towards output. In this way, after N round trips the overall dispersion that recursive CBGs impart is 2N times larger than the dispersion of each CBG.

To demonstrate this concept, we probe the system with pulses centered at 5 different wavelengths in the range of 1551 nm – 1559 nm, and measure the total time delay they experience. We note that since the pulse bandwidth is relatively narrow (on the order of 20 GHz), the impact of the waveguide dispersion on the pulse broadening is negligible. The results, obtained for different number of round trips N are shown in Fig. 4b. We measure a total delay of around 230 ps between 1551 nm and 1559 nm, corresponding to >28 ps/nm linear group delay achieved across the 30 nm wavelength range of the CBG bandgap (see supplementary information for simulation and experimental details). This amount of dispersion is sufficient to create integrated optical spectrum analyzers[48] and time-domain pulse shapers[36] for example.

**Reconfigurable differentiation by recursive asymmetric Mach-Zehnder interferometer**

One key advantage of the proposed recursive unit is its high degree of controllability. Here, we utilize our recursive approach to perform temporal derivatives up to fifth order on the input pulse train. This is accomplished by inserting an asymmetric Mach-Zehnder interferometer (AMZI) in the recursive loop (Fig. 4c). AMZIs have been used previously to perform 1st and 2nd derivative of electric fields of input pulses centered around its null transmission point[25]. (For details please see the supplementary information.) This approach, as well as the approach based on self-coupled optical waveguides[49], has been hard to scale to higher orders due to geometric variations between the fabricated devices that lead to different operating wavelengths and compromised performance. Furthermore, these approaches were not reconfigurable, and each device could perform one derivative only. In contrast, in our recursive circuit, the pulse is always processed using the same

AMZI, and the order of the derivative is simply controlled by changing the number of roundtrips that pulse makes (i.e. the Nth derivative corresponds to N round trips inside the loop). Here we demonstrate temporal derivatives from $0^{th}$ up to $5^{th}$ order (Fig. 4c, solid lines), and the results are in excellent agreement with theoretical predictions (Fig. 4c, dashed lines). For details of the comparison and numerical simulation, see supplementary information.

Finally, owing to the high tuning speed of our EO MZI switch, we also generate an arbitrary combination of different orders of differentiations from a single pulse. By operating the switch in a partially closed state, a sequence of increasing differential orders can be generated, each occupying a specific temporal duration Furthermore, the relative amplitudes of the derivatives of different order can be tuned arbitrarily by controlling the coupling strength between the waveguide and recursive loop (Fig. 4d). Different sequences are formed from successive pulses from the input pulse train. Inside them the amplitude of one specific differential order is deliberately set to 0 to show the extinction ratio. Normalized relative amplitudes of the derivatives are extracted and compared with ideal target and simulation. The device is capable of performing accurate reconfigurable differentiation with precise microwave control as supported by the agreement between simulation and experiment.

## Conclusion and Outlook

In summary, we proposed a novel on-chip electro-optic processing method based on a reconfigurable recursive circuit and performed a wide range of time-frequency domain operations. The key components, fast MZI switches and low loss optical waveguides, are implemented in TFLN photonics. We trapped and processed an optical packet as it makes up to 14 roundtrips, limited by the optical losses of the current device. With the state of art quality factor, which is around 30 million[12], and smaller loop length for shorter pulses, the estimate number of rounds can easily go up to 500. Introducing gain into the loop could further reduce the overall losses. We demonstrated three different functionalities using this concept: THz range frequency shifting of optical packets, large dispersion/delay, and reconfigurable high-order signal differentiation. Importantly, our approach not only achieves these functionalities in a tiny footprint, but also offers unmatched performance compared to conventional (e.g. waveguide or static cavity based) approaches. In future, extending this recursive approach to other functionalities, could allow for unprecedented temporal and spectral control of optical light (Fig. 1c). For example, narrow filters, which usually suffer from either limited extinction ratio provided by a single device or misalignment of center wavelengths of multiple devices due fabrication variations, could be made to achieve simultaneously a high extinction and narrow bandwidth. Furthermore, if combined with fast electrical feedback circuit, the circuit can excel in applications such as measurement-based quantum computation[50], where detection events can be used to trigger a change in the later rounds of operations.

**Data availability** The data that support the plots within this paper and other findings of this study are available from the corresponding author upon reasonable request.

**Code availability** The code used to produce the plots within this paper is available from the corresponding author upon reasonable request.

**Acknowledgements** We thank M. Zhang and A. Shams-Ansari for helpful discussions. This work is supported by National Science Foundation (EEC-1941583, ECCS-2407727, OMA-2137723, 2138068); Defense Advanced Research Projects Agency (HR0011-24-2-0360); Swiss National Science Foundation (grant 200020_192330/1); National Research Foundation of Korea; Amazon Web Services; DRS Daylight Solutions, Inc.; MagiQ Technologies: Naval Air Warfare Center Aircraft Division. A. C. acknowledges Rubicon postdoctoral fellowship from the Netherlands Organization for Scientific Research. Device fabrication was performed at the Harvard University Center for Nanoscale Systems. U. S. acknowledges the Postdoc.Mobility fellowship from the Swiss National Science Foundation.

**Competing interests:** M.L. is involved in developing lithium niobate technologies at HyperLight Corporation. The authors declare no other competing interests.

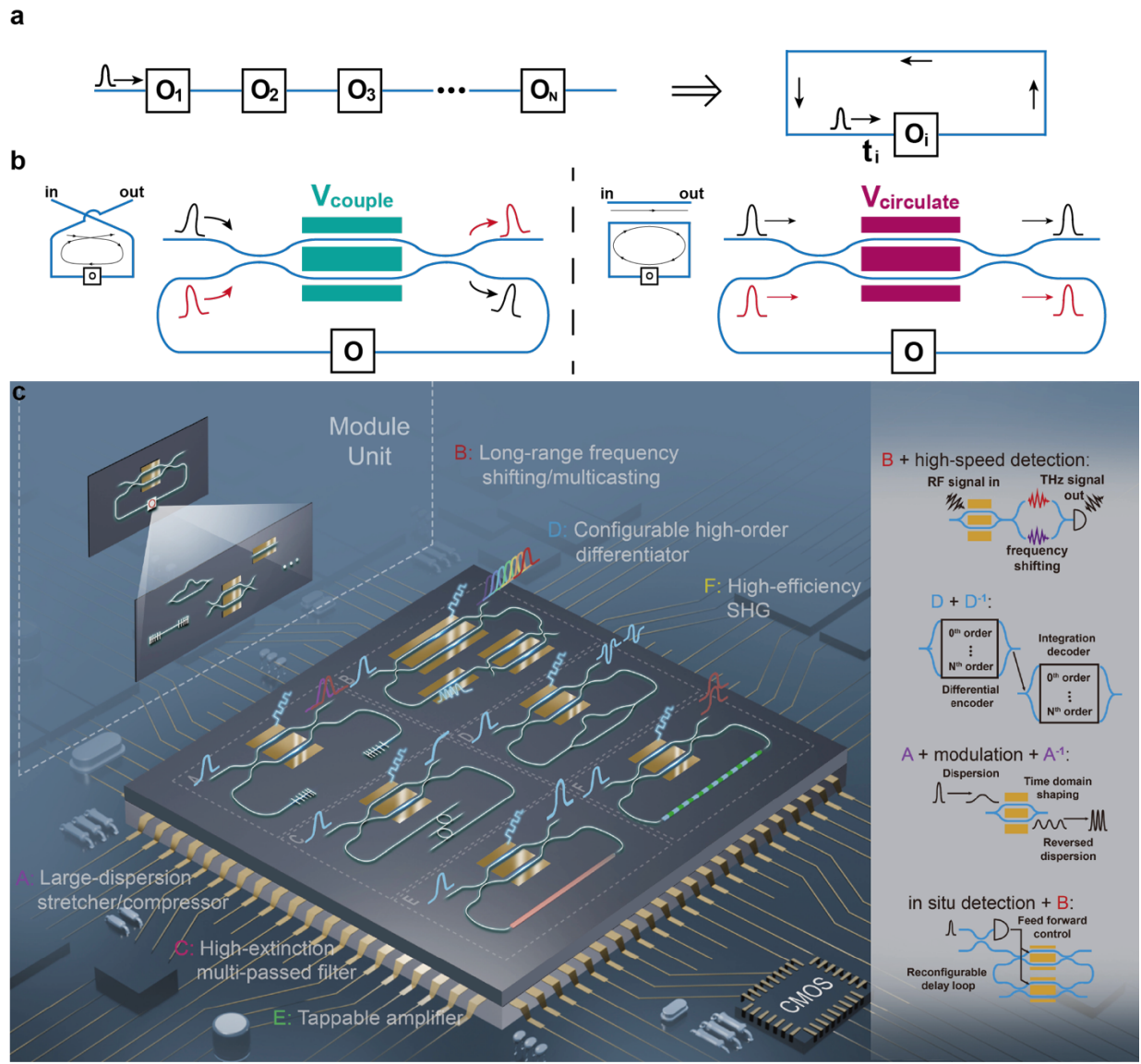

**Fig. 1 | Recursive processing scheme. a,** The operating principle of the recursive processing structure. In a conventional approach, optical pulse goes through several blocks, indicated as $O_1, O_2, .... O_N$, daisy-chained on the same waveguide, to perform desired processing task. Alternatively, the same processing task can be accomplished by keeping the pulse in a loop that consists of a reconfigurable block that can perform different functionalities at different points in time, as long as the block can be reconfigured on a time scale much shorter than a loop roundtrip time. In this way desired operations on an optical pulse can be effectively done by a single, ideally time-varying, functional block placed inside a loop. **b,** Working principle of the Mach-Zehnder Interferometer (MZI) switch-coupled resonator for circulating pulses. In coupled state, the pulses are guided into/out of the waveguide. In circulating state, pulses are kept inside the resonator and interact with the optical block of interest (O) multiple times. **c,** Concept of the proposed architecture implemented in TFLN platform. Top left: The circuit consists of a resonator with electric-optically controlled coupling rate with desired operation embedded inside. This configuration can be used for a wide range of applications by placing different optical blocks inside the loop, including phase modulator (PM), chirped Bragg gratings (CBG), asymmetric Mach-Zehnder interferometer (AMZI), amplitude modulator (AM). Middle: Examples of functionalities that could be accomplished. Right: Applications that are enabled by the modules, including THz signal synthesis, optical signal processing, time domain pulse shaping and feed forward control with detection.

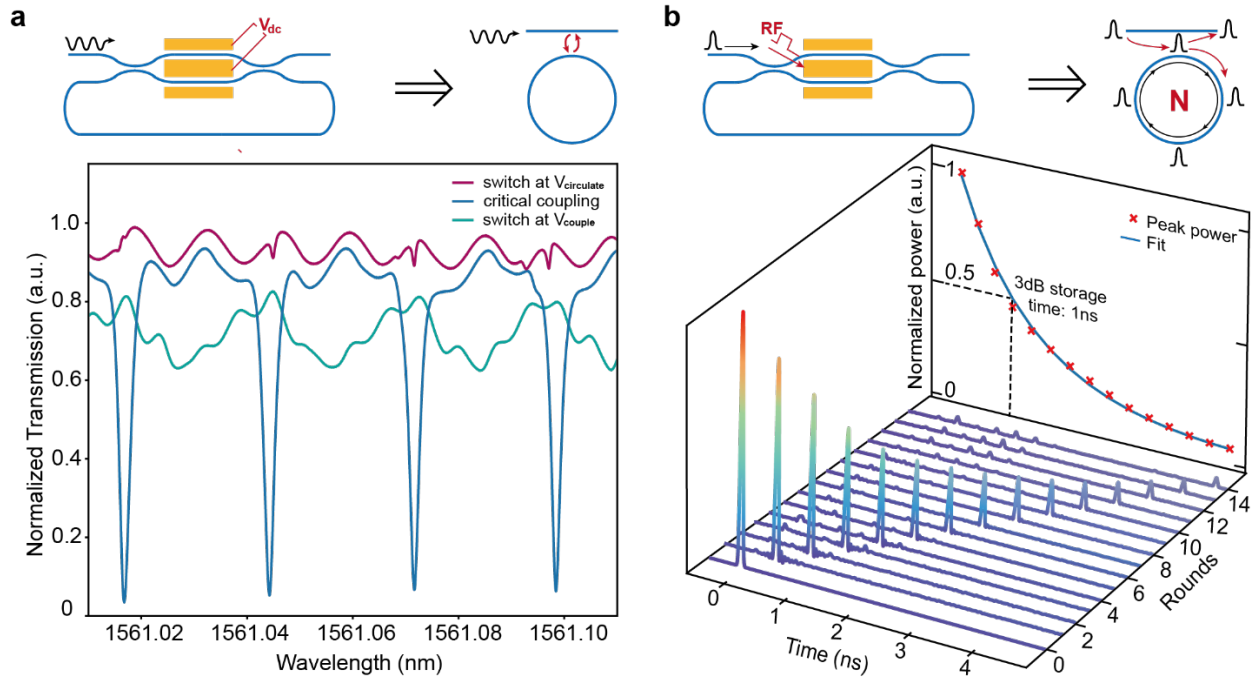

**Fig. 2 | Characterization of the reconfigurable recursive unit. a,** Transmission spectrum of the unit when different control voltages are applied to the switch thus controlling the coupling strength between the ring and the bus waveguide. Under-coupled (red), over-coupled (green) and critically coupled (blue) regime can be separately achieved. **b**, Characterization of pulse-trapping ability of the recursive unit under microwave control. In this case the switch is used to trap the pulse and to control the number of round trips that the pulse makes inside the loop/ resonator. Small peaks correspond to the light leaking outside the resonator when it is decoupled from the waveguide and are caused by a combination of finite extinction ratio of the MZI switch and variations in the voltage applied to the switch. The former is due to imperfect (not 100%) coupling/isolation of the loop since fabricated directional couplers do not have perfect 50:50 splitting. Inset: Fitting of peak pulse power to the number of rounds inside the loop which indicates -0.95 dB loss per round-trip.

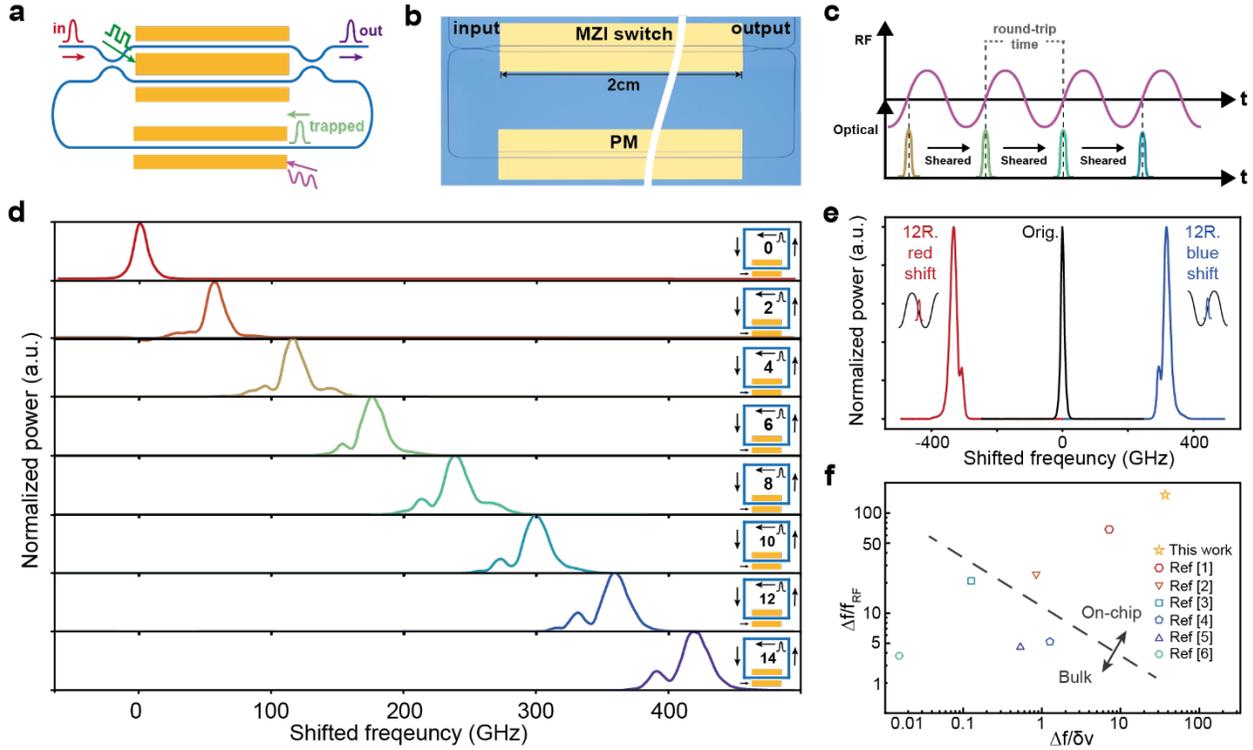

**Fig. 3 | Recursive unit for enhanced spectral shearing. a,** On-chip circuit scheme and **b,** Optical microscope image of the fabricated recursive unit for spectral shearing. Large spectral shifts for optical pulses can be achieved by placing a phase modulator (PM) inside the loop. **c,** Simplified scheme for shearing signal and incoming pulses. When the period and phase of the sinewave used to drive the PM is synchronized with the roundtrip time of the loop and the arrival time of the pulse (to match rising edge of the sinewave), large frequency up-shifts can be achieved. **d,** Spectra of the up-shifted pulses after 0-14 rounds of shearing, displayed in intervals of 2 rounds (round number denoted on the right). The peak power of each spectrum is normalized to unity. Side peaks are attributed to leakages from neighboring rounds due to non-ideal coupling condition. Broadening of the main peak is attributed to the walk-off between incoming pulse and the center of linear rising slope of the sinusoidal shearing wave. **e,** Frequency up (down) shift using rising (falling) slope of the sine wave from 12 rounds of shearing. Slightly reduced power was applied to PM to achieve better pulse shape. **f,** Comparison between this work and existing works on shearing in figures of merit $\frac{\Delta f}{f_{RF}}$ and $\frac{\Delta f}{FWHM}$.

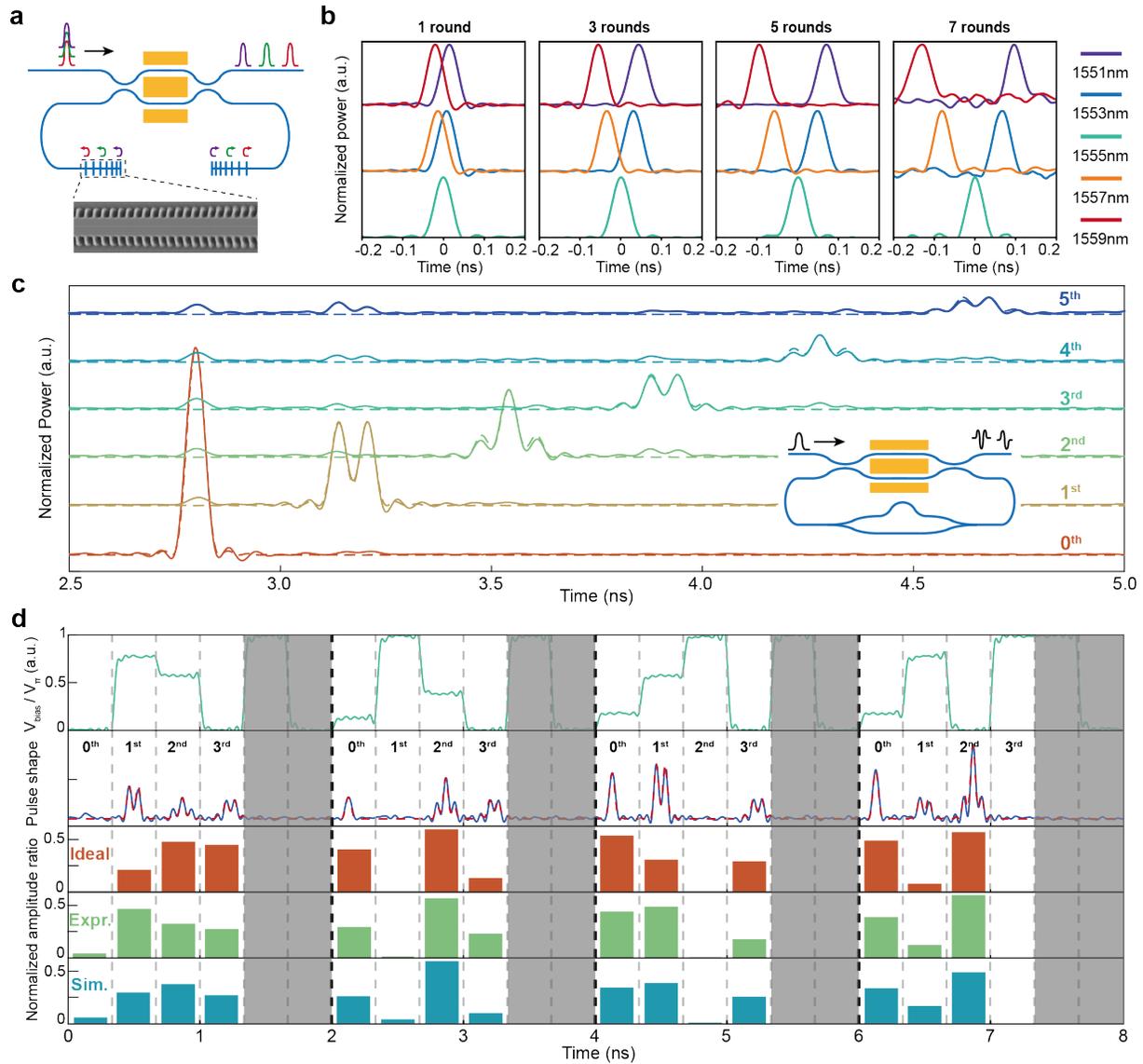

**Fig. 4 | Recursive units for temporal control of optical pulses. a,** Schematic of the recursive circuit used to achieve large dispersion with chirped Bragg gratings (CBG). When the switch is in circulate state, the pulse is trapped inside a linear Fabry-Perot cavity formed by the two CBGs and acquires additional delay upon each reflection. Inset: Scanning electron microscope (SEM) image of the fabricated CBG. **b,** Pulses with center wavelengths ranging linearly from 1551 nm to 1559 nm are injected into the recursive unit, and time-domain traces for 1-7 rounds of reflection is obtained. The pulse centered at 1555 nm is set to 0 to show the relative delay between different wavelengths. Around 30ps/nm linear group dispersion is achieved after 7-round operation of the chirped Bragg mirrors. **c,** $0^{th}$ to $5^{th}$ order temporal derivatives of the optical pulses can be obtained using imbalanced asymmetric MZI (AMZI) placed inside the recursive unit (inset) with different rounds of passing. Experiment and simulation results are shown with solid and dashed lines, respectively. (For details see supplementary information). **d,** Arbitrary sequences of temporal derivatives are generated by applying different voltages to the control switch during each roundtrip. The repetition rate of incoming pulse train is set to 0.5 GHz, enabling 6 time-bin channels where the first four are used to host $0^{th}$ to $3^{rd}$ order derivatives while the remaining two are left empty (Grey region). Green line: electrical control voltage; Blue/ Red dashed line: experimental/ simulated time-domain trace of optical pulse trains. Bar plots: ideal, experimental and simulated ratios which are encoded by the amplitudes of each order of differential signal.